\documentclass[12pt,preprint]{aastex}
\usepackage{emulateapj5,psfig,apjfonts}

\let\lsim=\la
\let\gsim=\ga

\newcommand{\EWobs}{{\rm EW}_{\rm obs}}
\newcommand\deltasigma{\delta_\Sigma}

\slugcomment{ApJ, accepted}

\shorttitle{Clustering of Lyman $\alpha$ Emitters at $z \simeq 5$}
\shortauthors{Shimasaku et al.}

\begin{document}


\title{Large Cosmic Variance in the Clustering Properties 
of Lyman $\alpha$ Emitters at $z\simeq 5$\altaffilmark{1}}


\author{K. Shimasaku   \altaffilmark{2,3},
  T. Hayashino  \altaffilmark{4},
  Y. Matsuda    \altaffilmark{4},
  M. Ouchi      \altaffilmark{2},
  K. Ohta       \altaffilmark{5}, \\
  S. Okamura    \altaffilmark{2,3}, 
  H. Tamura     \altaffilmark{4},
  T. Yamada     \altaffilmark{6},
  and R. Yamauchi    \altaffilmark{4}
}

\email{shimasaku@astron.s.u-tokyo.ac.jp}


\altaffiltext{1}{Based on data collected at Subaru Telescope, 
which is operated by the National Astronomical Observatory of Japan.}

\altaffiltext{2}{Department of Astronomy, School of Science,
        University of Tokyo, Tokyo 113-0033, Japan; 
shimasaku@astron.s.u-tokyo.ac.jp, 
ouchi@astron.s.u-tokyo.ac.jp,
okamura@astron.s.u-tokyo.ac.jp.}
\altaffiltext{3}{Research Center for the Early Universe, 
        School of Science,
        University of Tokyo, Tokyo 113-0033, Japan.}
\altaffiltext{4}{Research Center for Neutrino Science,
Graduate School of Science, Tohoku University,
Aramaki, Aoba, Sendai 980-8578, Japan; 
haya@awa.tohoku.ac.jp, 
matsuda@awa.tohoku.ac.jp,
tamura@awa.tohoku.ac.jp, 
yamauchi@awa.tohoku.ac.jp.}
\altaffiltext{5}{Department of Astronomy, Kyoto University,
Sakyo-ku, Kyoto 606-8502, Japan; 
ohta@kusastro.kyoto-u.ac.jp.}
\altaffiltext{6}{National Astronomical Observatory of Japan, 
Mitaka, Tokyo 181-8588, Japan; 
yamada@optik.mtk.nao.ac.jp.}


\begin{abstract}

We reported in a previous paper 
the discovery of large-scale structure of Lyman $\alpha$ 
emitters (LAEs) at $z=4.86\pm 0.03$ 
with a projected size 
of $20 h_{70}^{-1}$ Mpc $\times$ $50 h_{70}^{-1}$ Mpc 
in narrow-band data 
of a $25' \times 45'$ area of the Subaru Deep Field 
($\Omega_0=0.3, \lambda_0=0.7, H_0 = 70 h_{70}$ km s$^{-1}$ Mpc$^{-1}$).
However, the surveyed area, which corresponds to 
$55 h_{70}^{-1}$ Mpc $\times 100 h_{70}^{-1}$ Mpc, was
not large enough that we can conclude that we are seeing a typical 
distribution of $z\simeq 5$ LAEs.
In this Letter, we report the results of follow-up imaging 
of the same sky area using a new narrow-band filter 
(NB704, $\lambda_c=7046$\AA\ and FWHM$=100$\AA) to detect 
LAEs at $z=4.79$, i.e., LAEs lying closer to us by 
$39 h_{70}^{-1}$ Mpc on average than the $z=4.86$ objects.
We detect 51 LAEs at $z=4.79 \pm 0.04$ down to ${\rm NB704}=25.7$, 
and find that their sky distribution 
is quite different from the $z=4.86$ LAEs'.
The clustering of $z=4.79$ LAEs is very weak on any scales and 
there is no large-scale high-contrast structure.
The shape and the amplitude of the angular correlation function 
are thus largely different between the two samples.
These results demonstrate a large cosmic variance 
in the clustering properties of LAEs 
on scales of $\sim 50 h_{70}^{-1}$ Mpc.

\end{abstract}




%
%

\section{INTRODUCTION}
\label{sec:introduction}

Search for Lyman $\alpha$ emission of galaxies 
using a narrow-band filter is a powerful tool to detect 
high-$z$ faint galaxies. 
Indeed, many observations have successfully detected 
such Lyman $\alpha$ emitters (LAEs) from $z\sim2$ up to $z\simeq 6.6$ 
(e.g., 
Hu, Cowie, \& McMahon 1998; 
Pascarelle, Windhorst, \& Keel 1998; 
Campos et al. 1999; 
Hu, McMahon, \& Cowie 1999; 
Kudritzki et al. 2000; Rhoads et al. 2000;
Stiavelli et al. 2001; 
Ajiki et al. 2002; Hu et al. 2002; Venemans et al. 2002; 
Fynbo et al. 2003; 
Kodaira et al. 2003; 
Ouchi et al. 2003a; Shimasaku et al. 2003; 
see also Maier et al. 2003).
Narrow-band surveys can also map effectively 
large-scale distributions of LAEs  
(e.g., Steidel et al. 2000; Venemans et al. 2002; Hu et al. 2004).

We recently made a survey of LAEs at $z=4.86$ 
using the narrow-band filter NB711 ($\lambda_c=7126$\AA, 
FWHM=73\AA) in an area of $25' \times 45'$, and found 
a large-scale structure of LAEs 
of $20h_{70}^{-1}$ Mpc $\times$ $50h_{70}^{-1}$ Mpc
size (Shimasaku et al. 2003). 
This is the first discovery of large-scale structure in 
young universes, suggesting that the birth of large-scale structure 
is very early in the history of the universe 
and that LAEs are strongly biased against dark matter, 
since Cold Dark Matter (CDM) models predict 
that the density fluctuations of dark matter are very small 
on such large scales.
However, the size of the large-scale structure is nearly 
comparable to the size of the survey region 
($55h_{70}^{-1}$ Mpc $\times$ $100h_{70}^{-1}$ Mpc), 
and thus we cannot safely conclude that 
we are seeing a typical distribution of LAEs at $z\simeq 5$.
To address this issue, 
it is strongly needed to enlarge the survey volume.

Motivated by this, we made a followup imaging survey of LAEs 
at $z=4.79$, i.e., LAEs located closer to us by $39h_{70}^{-1}$ Mpc 
than those at $z=4.86$, in exactly the same sky area. 
Using these data, we examine differences in the sky distribution 
of LAEs between the two redshifts. 
We adopt $\Omega_0=0.3$, $\lambda_0=0.7$, 
and $H_0 = 70 h_{70}$ km s$^{-1}$ Mpc$^{-1}$.

\section{DATA}

We carried out deep imaging in the sky area of 
the Subaru Deep Field 
(SDF; centered at $(13^h 24^m 21.^s4, +27^\circ 29' 23'')$ 
[J2000.0]; Maihara et al. 2001) 
in a narrow-band filter centered at 7046 \AA, NB704,
with the prime-focus camera (Suprime-Cam; Miyazaki et al. 2002) 
on Subaru on 2003 February 2 and 3.
The FWHM of the NB704 filter is 100 \AA, 
giving a survey depth for LAEs along the sightline of 
$\Delta z = 0.08$ or, equivalently, $45 h_{70}^{-1}$ Mpc.
We observed two field-of-views (FoVs) of Suprime-Cam 
with a large overlap, the central FoV and the northern FoV, 
to obtain data for 
the same region ($25' \times 45'$) as imaged in the NB711 
in search for $z=4.86$ LAEs
\footnote{
The original $25' \times 45'$ area was determined as follows.
We first observed the central FoV in NB711, 
finding in a northern part
a large-scale overdense region of LAE candidates.
We then imaged the northern FoV with an overlap of $15'$ 
to trace a northern extension of the overdense region.
}.
For both filters, 
the variation in the response profile is small 
across the field of view even through the fast (F/1.86) 
prime-focus optics, with a full variation in $\lambda_c$ 
of 10\AA \hspace{2pt} (NB704) and 12\AA \hspace{2pt} (NB711).
This ensures a considerably uniform selection of LAEs 
with respect to redshift over the survey field.

Individual CCD data are reduced and
combined using IRAF and our own data reduction software 
(Yagi et al. 2002).
The seeing sizes of the final images are $0.''90$.
To detect LAEs, we combine broad-band $R$ and $i'$ data 
with the NB704 data. 
These $R$ and $i'$ data, taken in the spring of 2001, 
have been used to detect LAEs at $z=4.86$ combined with 
the NB711 data (Ouchi et al. 2003a; Shimasaku et al. 2003).
The total exposure time and the limiting magnitude 
($3\sigma$ on a $1.''8$ aperture) are: 
90 min and 27.1 mag ($R$), 
138 min and 26.9 mag ($i'$), and
198 min and 26.7 mag (NB704) for the central FoV; 
120 min and 27.5 mag ($R$), 
110 min and 27.3 mag ($i'$), and
216 min and 26.8 mag (NB704) for the northern FoV. 
All magnitudes are AB magnitudes.

Object detection and photometry are 
made using SExtractor version 2.1.6 (Bertin \& Arnouts 1996).
The NB704-band image is chosen to detect objects. 
We define the NB704-band limiting magnitude and the selection criteria 
for $z=4.79$ LAEs so that the lower limit to 
the Lyman $\alpha$ flux and the lower limit to 
the observed equivalent width of Lyman $\alpha$ emission ($\EWobs$) 
be nearly equal to those for the $z=4.86$ LAE sample 
in Shimasaku et al. (2003).
We set the limiting magnitude to be NB704=25.7, 
corresponding to 
$f({\rm Ly}\alpha) = 1.1 \times 10^{-17}$ erg s$^{-1}$ cm$^{-2}$ 
when the contribution from the continuum emission is negligible.
The total number of objects down to NB704$=25.7$ is 42,440.

\vspace{20pt}
\centerline{\psfig{file=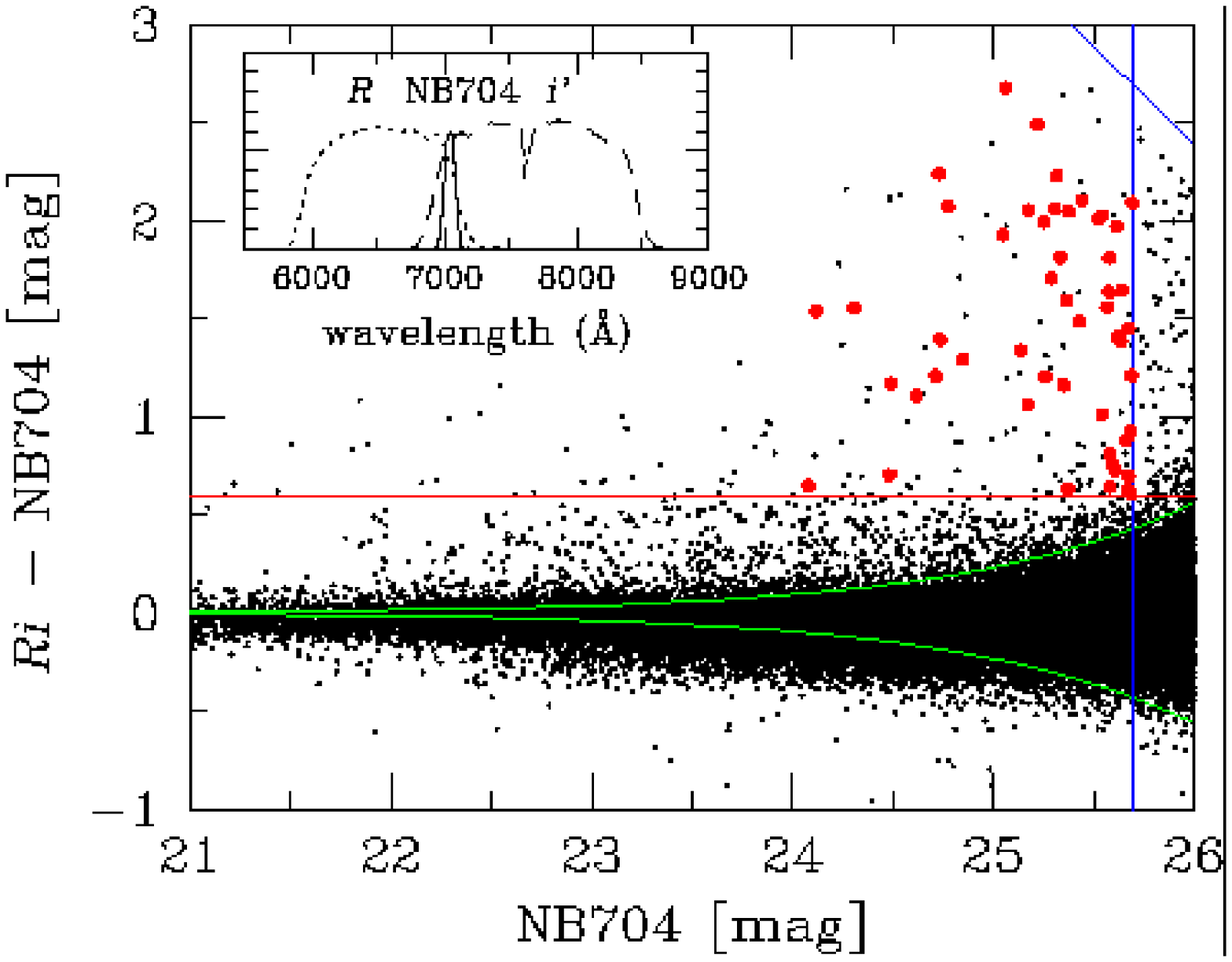,width=3.5in}}
\noindent{\scriptsize \addtolength{\baselineskip}{-3pt}
{\sc Fig.~1.} ---
Distribution of all the detected objects
in the $Ri - {\rm NB704}$
vs. NB704 plane, where $Ri \equiv (R + i)/2$.
The green lines indicate the distribution of $3\sigma$
errors in brightness for a source with a flat
($f_\nu$ = const) spectrum.
The blue vertical line shows the faint limit to our catalog,
and the blue diagonal line is for the detection limit of $Ri$
(average of the two FoV data).
The red line indicates our selection criterion on
$Ri - {\rm NB704}$ color.
The red filled circles are our photometrically selected
$z = 4.79$ LAEs that satisfy all the criteria (see text).
The small inserted panel plots the response functions of
the three filters (solid: NB704, dotted: $R$, dashed: $i'$).
}
\medskip

We apply the following three criteria to the detected objects 
to isolate LAEs at $z = 4.79$:
$Ri - {\rm NB704} > 0.6$, $R - i > 0.5$, and $i'-{\rm NB704}>0$, 
where $Ri \equiv (R+i')/2$.
Since the NB704 band measures fluxes between the $R$ and $i'$ bands, 
we define the off-band continuum flux of objects 
as $Ri \equiv (R + i)/2$. 
The first criterion, $Ri - {\rm NB704} > 0.6$, is set to select 
LAEs with $\EWobs \ge 80$ \AA.
The second and the third criteria are the same as those used
to construct the $z=4.86$ LAE sample.
The second criterion reduces contamination by foreground galaxies 
whose emission lines other than Lyman $\alpha$ 
happen to enter the NB704 band.
Figure 1 plots all objects with NB704$\le26$ 
in the $Ri-{\rm NB704}$ vs. NB704 plane.

The number of objects passing the above criteria is 51.
We have found that the contamination to the $z=4.86$ LAE sample 
from low-$z$ interlopers is about 20\% 
on the basis of spectroscopic observations and 
Monte Carlo simulations (Shimasaku et al. 2003). 
We expect a similar contamination rate for the $z=4.79$ LAE sample 
\footnote{ 
Although the NB704 data are deeper than the NB711 data, 
we have adopted considerably bright magnitudes 
for the limiting magnitudes of our samples.
In this case, contamination is determined mainly from the depths of 
$R$ and $i'$ data, which are common in the two LAE samples.
}.
We also expect that the completeness of the $z=4.79$ sample 
is close to that of the $z=4.86$ sample.

\vspace{10pt}
\centerline{\psfig{file=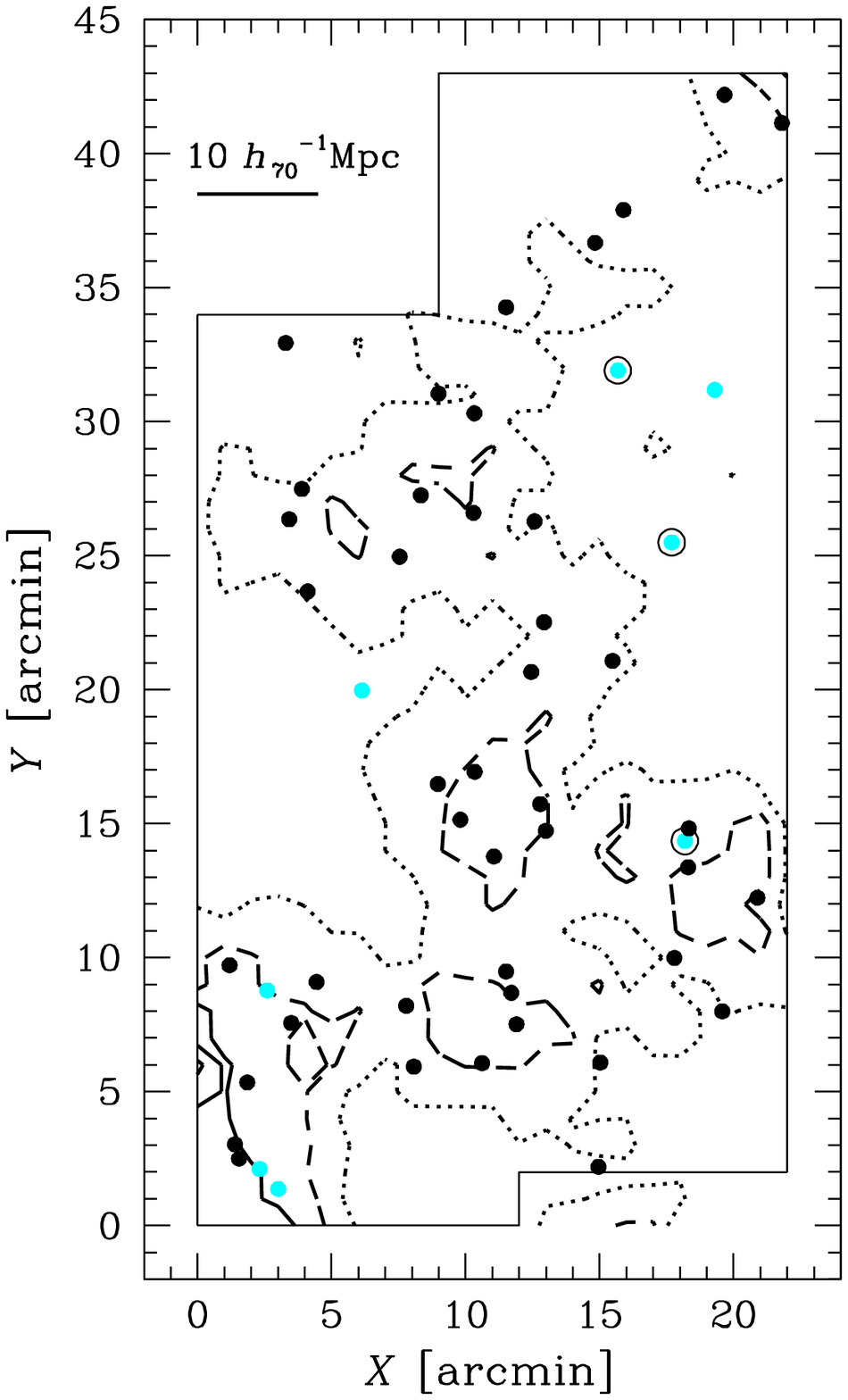,width=4.5in}}
\noindent{\scriptsize \addtolength{\baselineskip}{-3pt}
{\sc Fig.~2.} ---
Sky distribution of $z=4.79$ LAEs.
North is up and east is to the left.
Areas of relatively poor quality have been trimmed.
A bright star in the upper left corner, around which
detection of LAEs is impossible, has been masked.
LAE candidates are shown by circles.
Candidates common to the $z=4.86$ sample are marked by
open circles.
Objects with $-0.3<$NB704$-$NB711$<0.5$ are plotted in cyan.
The dotted, dashed, and solid lines correspond to contours of
$\deltasigma=0, 1$, and 2, respectively
(A top-hat smoothing of 8 Mpc radius
is made over the LAE distribution to compute
the local overdensity).
}

\section{RESULTS}

\subsection{Sky Distribution}

Figure 2 shows the sky distribution of 
photometrically selected 51 LAE candidates at $z=4.79$, 
while Figure 3 is a similar plot for 
43 $z=4.86$ LAE candidates.
For each sample, we estimate the local surface density of LAEs, 
$\Sigma(x,y)$, and compute the surface overdensity, 
$\deltasigma(x,y) \equiv [\Sigma(x,y)-\bar{\Sigma}]/\bar{\Sigma}$, 
where $\bar{\Sigma}$ is the mean surface density of LAEs 
($\bar{\Sigma}$ is defined for each sample).
The overdensity contours thus derived are drawn in each figure.
Three objects are found to be common to the two 
samples (open circles in Figs. 2 and 3).

We point out two large differences in the distribution of LAEs 
in the two samples.
First, the clustering of $z=4.79$ LAEs is much weaker than 
that of $z=4.86$ LAEs. 
In Fig. 2, no clear large-scale inhomogeneity is found, 
although the overdensity seems to become higher with decreasing 
$Y$.
In contrast, a large-scale concentration of LAEs 
and large {\lq}void{\rq} regions are seen in Fig.3.

\vspace{10pt}
\centerline{\psfig{file=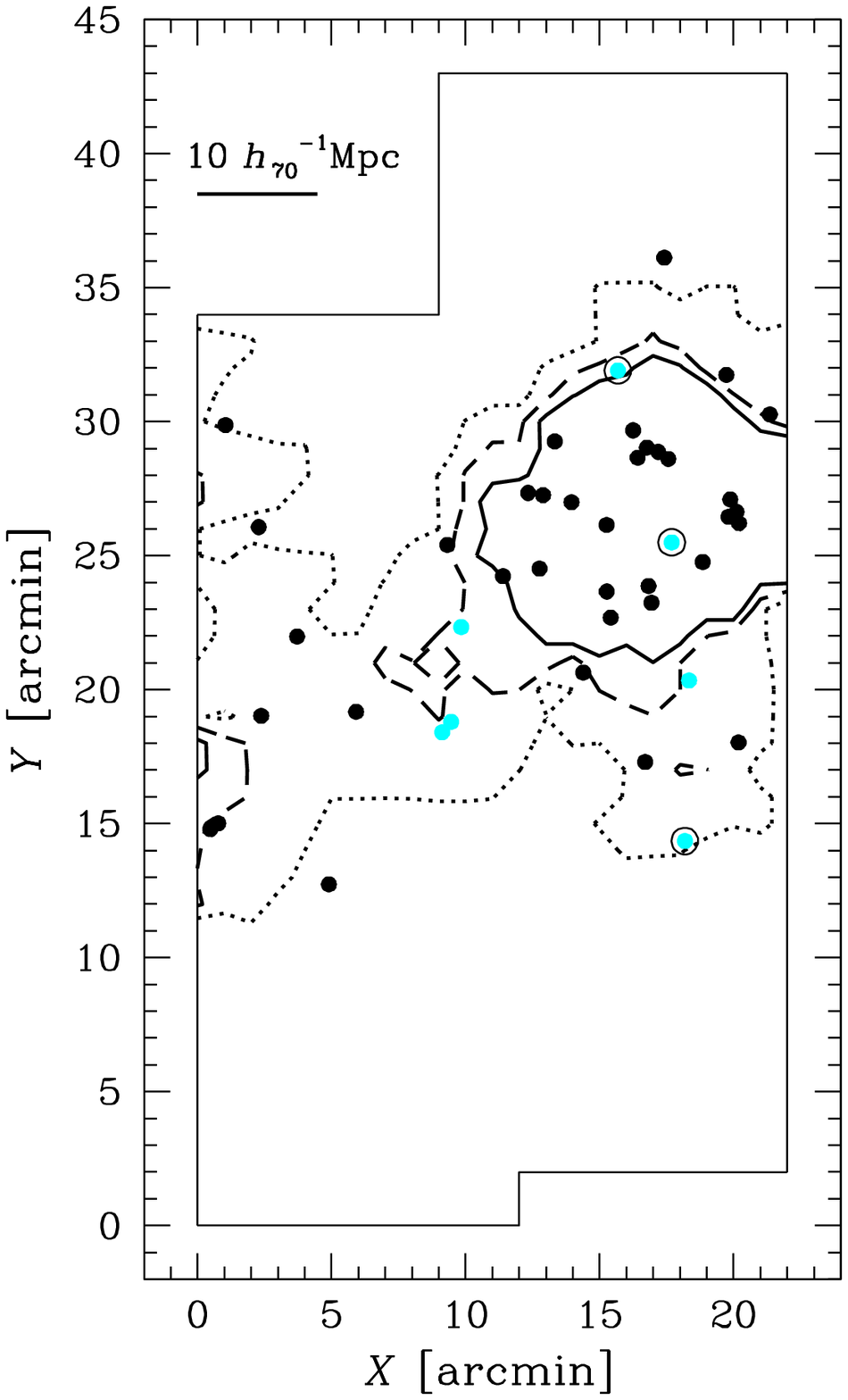,width=4.5in}}
\noindent{\scriptsize \addtolength{\baselineskip}{-3pt}
{\sc Fig.~3.} ---
Same as Fig.2, but for $z=4.86$ LAEs.
Candidates common to the $z=4.79$ sample are marked by
open circles.
}
\medskip

Second, no concentration of $z=4.79$ LAEs is found 
in the large-scale structure of $z=4.86$ LAEs 
at $Y\simeq 20-30$.
We identify objects which are likely to be at $z \sim 4.83$ 
from NB704$-$NB711 color ($-0.3<$NB704$-$NB711$<0.5$; 
12 objects in total from the two samples), 
and plot them in cyan in Figs.2 and 3.
We find in Fig.3 that the objects plotted in cyan are less 
concentrated on the sky than the others. 
We also note that the sky distribution of 
the 12 cyan objects seems to be in the middle 
of the distributions of
the $z=4.86$ LAEs and the $z=4.79$ LAEs.
These findings indicate that the line-of-sight extension of 
the $z=4.86$ large-scale structure toward lower redshift 
is not larger than $\sim 20 h_{70}^{-1}$ Mpc. 

\subsection{Angular Correlation Functions}

We calculate the angular correlation function (ACF), 
$\omega$($\theta$), for the two samples, 
using the estimator defined by Landy \& Szalay (1993), 
where $\theta$ is the angular separation.
The results are plotted in Figure 4.
The ACF for $z=4.86$ LAEs has large positive values 
at $\theta \lsim 8'$, 
while the amplitude of the ACF for $z=4.79$ LAEs is $\simeq 0$ 
at any angular separation.

\vspace{20pt}
\centerline{\psfig{file=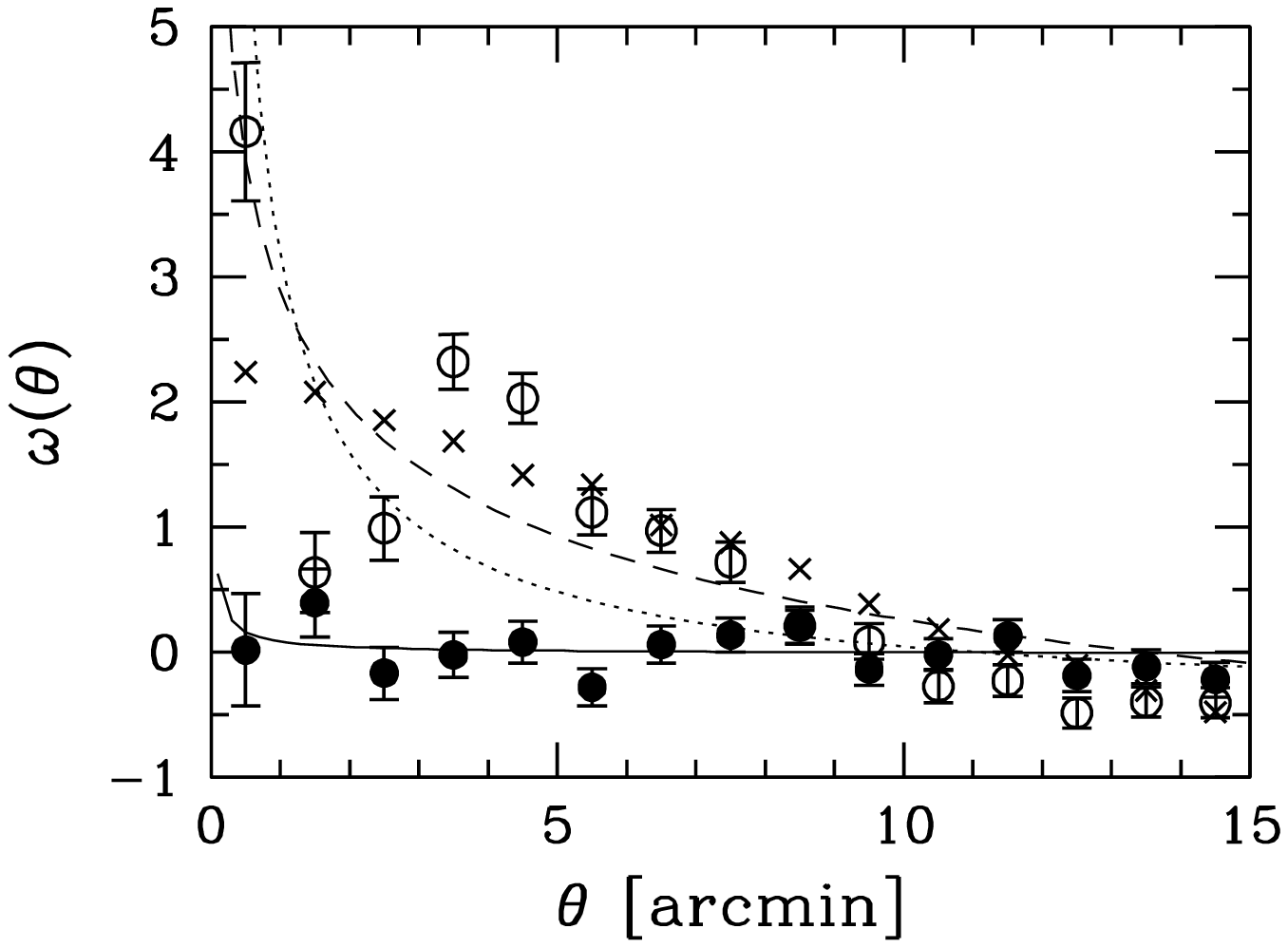,width=3.3in}}
\noindent{\scriptsize \addtolength{\baselineskip}{-3pt}
{\sc Fig.~4.} ---
Angular correlation functions (ACFs) of LAEs at $z=4.79$
(filled circles) and $z=4.86$ (open circles).
Errorbars correspond to the $1\sigma$ errors.
The solid line indicates the best fit of
$A_\omega (\theta^{-0.8} - C)$ to the $z=4.79$ data.
The dotted and dashed lines represent, respectively,
the best fit of $A_\omega (\theta^{-0.8} - C)$ and
$A_\omega (\theta^{-\delta} - C)$ to the $z=4.86$ data;
we find the best-fit $\delta$ to be 0.2.
The crosses indicate the ACF obtained if all LAEs lie in
the large-scale structure and their distribution in the
large-scale structure is uniform.
}
\medskip

The ACF of galaxies, 
including Lyman-break galaxies (LBGs), is approximated 
fairly well by a power-law function, $\theta^{-0.8}$ 
(for LBGs, see, e.g., Foucaud et al. 2003; Ouchi et al. 2003b).
We thus fit $\omega(\theta) = A_\omega (\theta^{-0.8} - C)$ 
to our LAE data 
to examine whether this function is also a good fit to the LAE data 
and to compare $A_\omega$ between the two samples, 
where $C$ is the integral constraint 
($C=0.15$ arcmin$^{-0.8}$).
We obtain $A_\omega = 0.10 \pm 0.21$ arcmin$^{0.8}$ 
and $\chi^2/N=1.1$ for $z=4.79$ LAEs. 
For $z=4.86$ LAEs, we obtain 
$A_\omega = 3.7 \pm 0.3$ arcmin$^{0.8}$, 
about 40 times higher than that for $z=4.79$ LAEs
\footnote{
Since the FWHM of the NB704 filter is wider than that of NB711 
by about 30\%, the ACF will be lower for $z=4.79$ LAEs 
even when the intrinsic, spatial correlation 
is the same. 
However, the difference in $A_\omega$ due to this effect 
is only about 15\%.
}, 
and find that the fit to the $z=4.86$ data is extremely 
poor ($\chi^2/N=14$). 
If we fit $A_\omega (\theta^{-\delta}-C)$ to the $z=4.86$ data, 
with $\delta$ being another free parameter, then we find 
a very flat power-law slope, $\delta=0.2$, 
although the fit with this slope is still unsatisfactory 
($\chi^2/N=10$).
For $z=4.79$ LAEs we obtain $\delta=0.9$.

The ACF of $z=4.86$ LAEs does not increase 
like $\theta^{-\delta}$, 
but remains more or less flat around $\omega \sim 1-2$ at $\lsim 8'$ 
except for the data point at $0.'5$.
Fig.3 shows that most of the $z=4.86$ LAEs belong to 
the large-scale structure  
which occupies about $30\%$ of the survey area.
For such an extreme distribution, it is shown that 
the ACF has high values, $\omega \sim 1$ -- $2$, 
at $\theta \lsim 8'$, 
even when we artificially re-distribute LAEs uniformly 
in the large-scale structure (crosses in Fig.4).
Thus, the high, relatively constant amplitudes of the observed ACF 
are largely due to the presence of the large-scale structure 
and the large void regions.

\subsection{Number Density}

We estimate the spatial number density of LAEs 
to be $(2.7 \pm 0.4) \times 10^{-4} h_{70}^{3}$ 
Mpc$^{-3}$ for the $z=4.79$ sample 
and $(3.1 \pm 0.5) \times 10^{-4} h_{70}^{3}$ 
Mpc$^{-3}$ for the $z=4.86$ sample.
In these calculations, we adopt a survey volume of 
$1.9 \times 10^5 h_{70}^{-3}$ Mpc$^3$ ($z=4.79$ sample) and 
$1.4 \times 10^5 h_{70}^{-3}$ Mpc$^3$ ($z=4.86$).
We have not applied completeness correction here.
The difference in the number density between 
the two samples is found to be within the $1 \sigma$ levels, 
although the $z=4.86$ LAEs might be slightly more numerous.

The rms fluctuation of mass overdensity at $z=4.8$ 
on a sphere with a volume equivalent to the average of 
our two survey volumes is calculated to be $\simeq 0.08$ 
for $\sigma_8=0.9$, 
where $\sigma_8$ is the present-day rms fluctuation 
of mass overdensity on a sphere of $8 h_{100}^{-1}$ Mpc radius 
($H_0=100 h_{100}$ km s$^{-1}$ Mpc$^{-1}$).
This implies that the number density of LAEs 
found in a volume similar to our average survey volume 
will fluctuate roughly by $\sim 0.08 b_g$ (if Poisson noise is 
not included), where $b_g$ is the bias parameter.
The observed small fluctuation would prefer relatively small
$b_g$ values.

\subsection{Possible Effects of Velocity Structures}

In the analyses above, we have not considered the possibility 
that our LAEs form structures along redshift 
({\lq}velocity structures{\rq}).
A recent spectroscopic observation of 
a wide-field LAE sample at $z\simeq 5.7$ has revealed distinct velocity 
structures with a depth of $\Delta z \sim 0.05$ 
(Hu et al. 2004; see also Fynbo et al. 2003 for a similar finding
on QSO-absorber fields at $z\simeq3$). 
LAEs in our samples might have such structures.
Multiple velocity structures within the filter bandpass 
make clustering analysis complicated.
If, for instance, our $z=4.79$ LAEs 
are divided into a few velocity structures, 
the true clustering amplitude will be higher than 
inferred from the apparent sky distribution 
(and from the ACF measurement).
On the other hand, the large-scale structure 
found at $z=4.86$, elongated along right ascension, 
may be an edge-on view of 
a sheet-like structure extending along the line-of-sight.
Velocity structures can also give an uncertainty 
to the estimates of number density made in \S 3.3.
Spectroscopic data are needed for a significant fraction 
of our samples to examine such possibilities.

\section{DISCUSSION}

Hamana et al. (2004) found that the observed ACF 
of $z=4.86$ LAEs given in Ouchi et al. (2003a)  
cannot be reproduced by a simple halo 
model which assumes LAEs to be associated with dark haloes, 
because of too strong the observed correlation on scales $ \gsim 2'$.
We find that our $z=4.86$ LAEs have large-scale 
structure and that their ACF is high and not fit by a power law, 
while the clustering of $z=4.79$ LAEs is very weak.
These results may suggest that 
the clustering of LAEs is typically weak, 
possibly tracing the dark-matter distribution, 
and that we happened to observe an unusual region 
in the $z=4.86$ universe where LAEs form a large, coherent 
structure of a size of $\sim 50 h_{70}^{-1}$ Mpc.
Conversely, if it turns out, from a larger survey, 
that the $z=4.86$ region we observed 
is relatively common in high-$z$ universes, this will suggest that 
LAEs and LBGs are separate populations 
in terms of clustering properties, 
since the clustering of LBGs has been found to be 
approximated well by halo models (Hamana et al. 2004).
Detailed modeling of the clustering of LAEs based on 
an enlarged sample will give important hints on the nature of LAEs.

Although we did not find in our two LAE samples 
a clear difference in the number density, 
a large field-to-field variance in the clustering of LAEs, 
including velocity structures, 
can influence 
measurements of the number density 
of LAEs based on narrow-band surveys, 
especially if the survey volumes are smaller than ours 
and if $b_g$ is much larger than unity; 
the $b_g$ value derived from the clustering of the $z=4.86$ LAEs is 
as large as $\sim 10$.
Shallower surveys will suffer from larger variances, 
since LAEs with brighter Lyman 
$\alpha$ luminosities tend to be clustered more strongly 
(Ouchi et al. 2003a).
For instance, 
Ajiki et al. (2003) found that 
the number density of $z=5.7$ LAEs in their sample 
of $2 \times 10^5 h_{70}^{-3}$ Mpc$^3$ 
is three times higher than that estimated 
by Rhoads \& Malhotra (2001) based on a similar survey volume.

To summarize, our observations show that 
it is necessary to survey a much larger volume than ours 
in order to derive the average clustering properties of LAEs.


\acknowledgments
We thank the anonymous referee for the useful comments.
M. O. acknowledges support from the Japan Society for the
Promotion of Science (JSPS) through JSPS Research Fellowships
for Young Scientists.




%
%

\end{document}